\documentclass[a4paper,
              ]{jacow}
              
\makeatletter%
	\ifboolexpr{bool{xetex}}
	 {\renewcommand{\Gin@extensions}{.pdf,%
	                    .png,.jpg,.bmp,.pict,.tif,.psd,.mac,.sga,.tga,.gif,%
	                    .eps,.ps,%
	                    }}{}
\makeatother

\usepackage[utf8]{inputenc}
\usepackage{amsmath}
\usepackage{breqn}
\usepackage{hyperref}
\usepackage{gensymb}
\usepackage{amsfonts}
\usepackage{amssymb}
\usepackage{graphicx}
\usepackage{lmodern}
\usepackage{mathtools}
\usepackage{siunitx}

\usepackage[USenglish]{babel}			 

\usepackage[final]{pdfpages}
\usepackage{multirow}
\usepackage{ragged2e}
\usepackage{lipsum}

\newcommand\blfootnote[1]{%
  \begingroup
  \renewcommand\thefootnote{}\footnote{#1}%
  \addtocounter{footnote}{-1}%
  \endgroup
}

\ifboolexpr{bool{jacowbiblatex}}%
 {%
  \addbibresource{jacow-test.bib}
  \addbibresource{biblatex-examples.bib}
 }{}
\author{A. Bellandi\thanks{andrea.bellandi@desy.de}, J. Branlard, A. Nawaz, R. Rybaniec, H. Schlarb, C. Schmidt,\\ Deutsches Elektronen-Synchrotron, Hamburg, 22607, Germany\\
W. Cichalewski, TUL, Łódz, 90-924, Poland}

\title{Simulation of microphonic effects in high Q\textsubscript{L} TESLA cavities }

\DeclarePairedDelimiter\abs{\lvert}{\rvert}%

\begin{document}
\maketitle

\begin{abstract}
\blfootnote{The authors of this work grant the arXiv.org and LLRF Workshop's International Organizing Committee a non-exclusive and irrevocable license to distribute the article, and certify that they have the right to grant this license.}
This document describes a new package to compute high performance simulations of a module of superconducting accelerating cavities from the LLRF controller perspective. The reason to make a dedicated C++/Python package is to simulate all the effects that arise during Continuous Wave (CW) operations at different timescales to speed-up the LLRF controller design. In particular the speed of the sampling rate of the ADCs used in a LLRF control system (some MHz) are $10^4$ - $10^5$ times faster than typical mechanical resonances and microphonics frequencies. 
\end{abstract}

\section{Introduction}
European X-Ray Free-Electron Laser (E-XFEL) and Free-electron Laser in Hamburg (FLASH) are the two main superconductive LINACs based on TESLA\cite{aune2000} cavities  at DESY and they are used to produce short-wave X-ray laser light through the FEL process. These machines are currently used in pulse mode and they produce a burst of short spaced bunches every 10 Hz. Because of the interest of relaxing the spacing between bunches there are proposals to turn XFEL and FLASH in Continuous Wave (CW) machines\cite{sekutowicz2013}. In such machines the accelerating gradient is constant and an arbitrary long train of particle can be accelerated. In order to do that the loaded quality factor (Q\textsubscript{L}) of the cavities has to be increased to keep the power and thermal requirements within reasonable  limits. There is also the need to keep the error on amplitude and phase of the RF field below 0.01 \% and $0.01\degree$\cite{branlard2012}.

At the moment tests for CW operations on cavity modules are performed at DESY, but varying QL above the nominal tuning range is a time consuming process and verifying the controller behaviour under beam loading is impossible at our current test stand because of the lack of an injector.

A code to simulate the cavity module with different Q\textsubscript{L} and beam loading  can be useful to speed-up the optimization of the controller.

\section{Simulation package}

The package code has some requirements in order to obtain: low simulation time, accuracy of the simulation and an easily modifiable LLRF controller code. Because of this it was decided to make the core simulator in C++ and to expose a public API in Python to implement the LLRF controller.

\begin{figure}
\centering
\includegraphics[width=174pt]{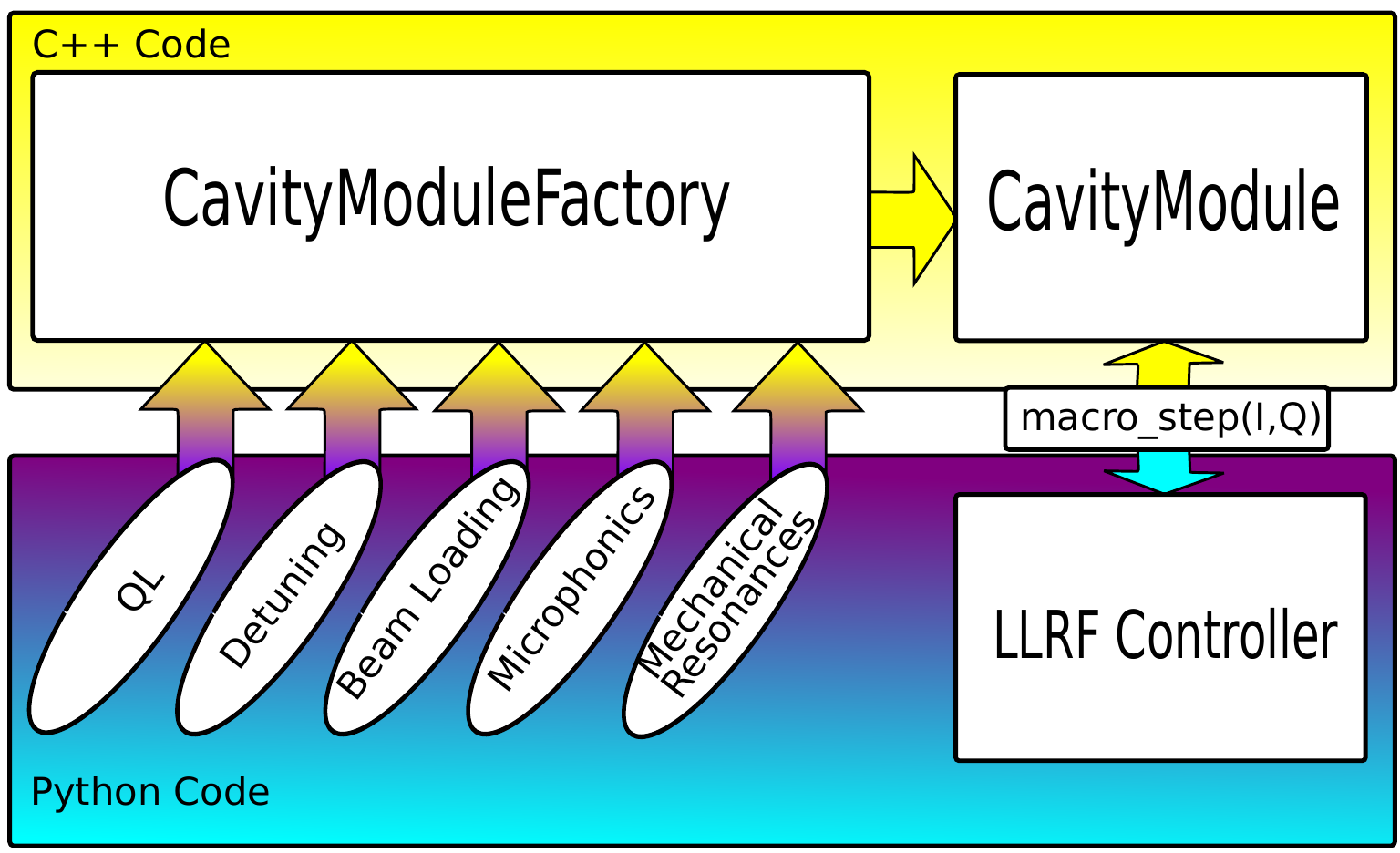}
\caption{From the Python side the properties of a 'Factory' object are set. Then a 'Module' object is created from the factory and used to simulate a cavity module. }
\end{figure}

The package is able to simulate a multi-cavity module with arbitrary Lorentz Force Detuning (LFD) mechanical resonances and microphonic sources\cite{liepe2001}. Beam loading effects can also be added.
The resulting system can be described with a canonical form nonlinear ODE system of equations:

\begin{equation}
\begin{split}
\dot{V}_I = \omega_{1/2}  V_I + (2 \pi k_{s}\abs{V}^2-\Delta\omega_{me}(t)-\Delta\omega_{mi}(t))V_Q+\\\frac{R}{Q}\omega_0I_{If}(t) + I_{Ib}(t)
\end{split}
\end{equation}
\begin{equation}
\begin{split}
\dot{V}_Q = \omega_{1/2}V_Q-(2 \pi k_{s}\abs{V}^2-\Delta
\omega_{me}(t)-\Delta\omega_{mi}(t))V_I+\\\frac{R}{Q}\omega_0I_{Qf}(t) + I_{Qb}(t)
\end{split}
\end{equation}
for the evolution of electrical voltage inside the cavity, whereas for mechanical oscillations:
\begin{equation}
\begin{split}	
\Delta\omega_{me}(t) = \sum^N_{i=1}\Delta\omega_i(t)
\end{split}
\end{equation}
\begin{equation}
\begin{split}
\ddot{\Delta\omega_j} = -\frac{(\omega^{me}_0)_j}{Q^{me}_j}\dot{\Delta\omega}_j -(\omega^{me}_0)^2_j \Delta\omega_j-\\2\pi(\omega^{me}_0)^2_jk_j\abs{V}^2
\end{split}
\end{equation}
\begin{equation}
\begin{split}
\Delta\omega_{mi}(t) = \sum^M_{i=1} (\Delta\omega_{a})_i\sin{((\omega_{mi})_it} + (\phi_{mi})_i)
\end{split}
\end{equation}
\begin{equation*}	
j = 1 .. N
\end{equation*}

are used, where:
\begin{itemize}
\item{$V_I$,$V_Q$,$I_{If}$,$I_{Qf}$,$I_{Ib}$ and $I_{Qb}$ are the IQ components of the accelerating voltage, the driving current and the beam current}
\item{$\omega_{1/2}$ is the cavity half-badwidth}
\item{$k_s$ is the coupling factor of static LFD}
\item{$\Delta\omega_{me}$ is the mechanical modes detuning, calculated as the sum of $N$ contributions $\Delta\omega_i$. Each contribution is produced by a mechanical mode with resonance frequency $(\omega_0^{me})_j$, quality factor  $Q_j^{me}$ and coupling $k^{me}_j$} 
\item{$\Delta\omega_{mi}$ is the microphonic contribution to the detuning, calculated as the sum of $M$ sinusoidal signals with amplitude $(\Delta\omega_a)_i$, frequency $(\omega_{mi})_i$ and phase $(\phi_{mi})_i$}
\end{itemize}

This system of equations can be used in conjunction of Boost C++ Odeint package \cite{ahnert2011} to make a step simulator. A comparison between a previous pure Python simulator was done in order to prove the correctness of the calculations.

\section{Comparison between simulated proportional controllers}

\begin{figure}
\centering
\includegraphics[scale=0.39]{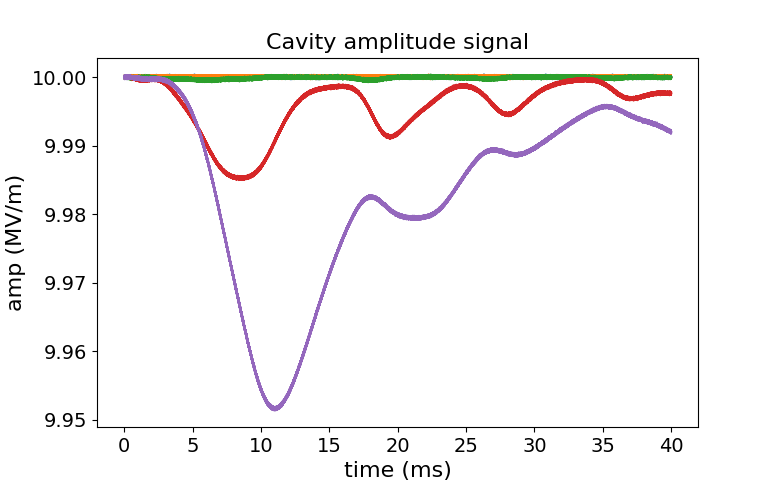}
\includegraphics[scale=0.39]{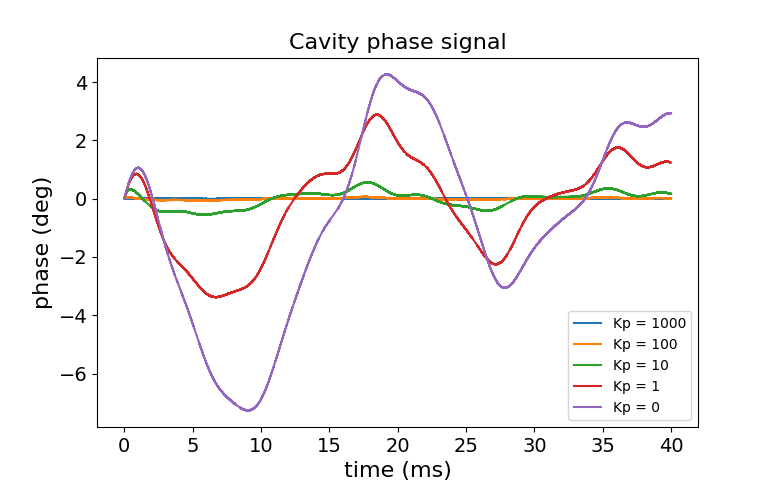}
\caption{Amplitude and phase evolution of the cavity gradient at different proportional gain values}
\label{fig:amppha}
\end{figure}

A first test was done simulating a single superconducting cavity with a Q\textsubscript{L} $= 2\cdot 10^7 $ and an accelerating gradient of 10 MV/m. For the mechanical resonances the HOBICAT\cite{neumann2008} parameters were used. For the microphonics two frequencies measured at DESY in CW tests were used:

\begin{itemize}
\item{$f_{micro1}$: 49 Hz with 5.0 Hz amplitude detuning}
\item{$f_{micro2}$: 31 Hz with 1.25 Hz amplitude detuning}
\end{itemize} 

A loop delay of $\SI{1.55}{\micro\second}$ was also simulated using a FIFO queue.

As shown in Fig \ref{fig:amppha}, increasing the gain value from $K_p=1$ to $K_p=1000$ reduces the signal error. Increasing the value to more than $K_p=3000$ resulted in instabilities of the loop.

\begin{figure}
\centering
\includegraphics[scale=0.31]{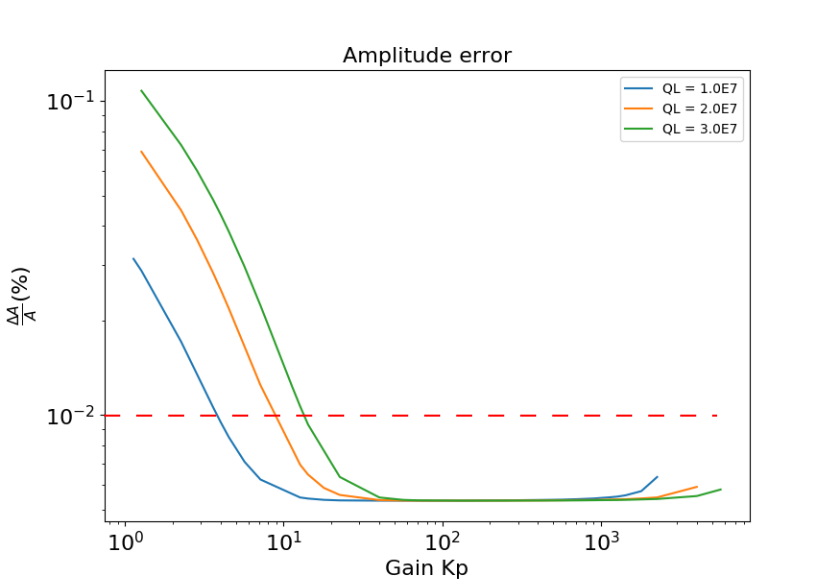}
\includegraphics[scale=0.31]{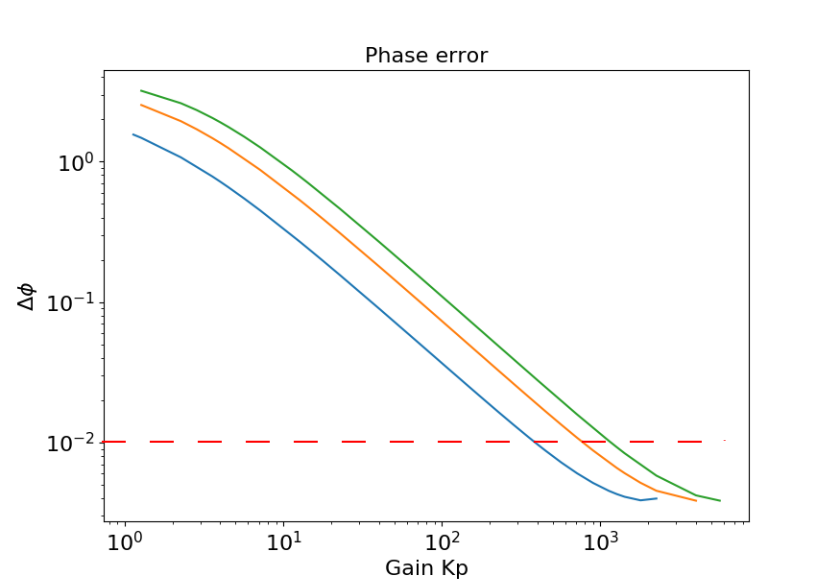}
\caption{Amplitude and phase RMS error of the cavity at different proportional gain values}
\label{fig:ampphaerr}
\end{figure}

Simulations were also performed at different values of Q\textsubscript{L} as shown in Fig \ref{fig:ampphaerr}. The results show a constant amplitude error of 0.005 \% $\frac{\Delta A}{A}$ after around $K_p > 50$ and a decreasing phase error became less than $0.01\degree$ after $K_p > 10^3$
\section{Conclusions}
In this document a new package for performing simulations of superconductive cavity modules from the LLRF perspective was presented. The results for preliminary studies are reasonable in respect to experimental data of the HOBICAT prototype. Further improvements of the code will be necessary. In particular differents loop algorithms (PI, MIMO) will be compared to find the best one for CW operations. The possibility to simulate piezo operations, needed to achieve low detuning in high Q\textsubscript{L} scenarios, will be added. Another improvement is to add the possibility to set multiple bunch trains to simulate multibeam delivery.

\end{document}